\documentclass[doublecol]{epl2} 

\usepackage[english]{babel}

\title{Geometrical view of quantum entanglement}

\author{A. Ram{\v s}ak\inst{1,2} }

\shortauthor{A. Ram\v sak}

\institute{                    
  \inst{1} Faculty of Mathematics and Physics, University of
  Ljubljana, Ljubljana, Slovenia\\
  \inst{2} Jo{\v z}ef Stefan Institute, Ljubljana, Slovenia
}

 
\abstract{Although a precise description of microscopic physical problems requires a full quantum mechanical treatment, physical quantities are generally discussed in terms of classical variables. One exception is quantum entanglement which apparently has no classical counterpart.
We demonstrate here how quantum entanglement may be within the de Broglie-Bohm interpretation of quantum mechanics visualized in geometrical terms, giving new insight into this mysterious phenomenon and a language to describe it. On the basis of our analysis of the dynamics of a pair of qubits, quantum entanglement is linked to concurrent motion of angular momenta in the Bohmian space of hidden variables and to the average angle between these momenta.}

\begin{document}

\maketitle

{{\bf Introduction.} --}
The recent explosion in activity related to entanglement is a consequence of a growing realization of its importance as a vital resource in quantum information through quantum teleportation\cite{tele}, quantum cryptographic key distribution\cite{key} and quantum computation\cite{compute,vedral} as well as its potential for enhanced quantum sensing through the ‘engineering’ of highly entangled quantum states, beating the usual quantum limit\cite{limit}. Despite this utility, the concept of entanglement remains mysterious and paradoxical within the usual quantum treatment and remains a focal point for discussing the foundations of quantum phenomena and related questions of non-locality\cite{buchanan}, with renewed impetus from the recent developments in quantum information and quantum technology. 

The connection between the classical and quantum approaches is usually made through the correspondence principle although an alternative description based on the historic work of de Broglie\cite{debroglie}, and  later developed by Bohm\cite{bohm1} gives equivalent results provided an effective non-local potential is introduced along with hidden variables. This approach has had a resurgence in interest recently re-emphasizing the virtues of the ontological description it provides\cite{valentini}.

With the development of the Bohm causal formulation of quantum mechanics\cite{bohm1}, first for single spinless particles and later for many-body fermionic or bosonic systems including fields\cite{holland00,durr}, it was shown that  the Bohmian approach is isomorphic with usual quantum mechanics regarding
observable predictions, but it additionally gives an interpretation of internal motion of particles in terms of classical coordinates, velocities and angular momenta. Although these (hidden) variables are not measurable, the approach is able to analyse individual processes in a way which goes beyond the standard Bohr interpretation of, for example, double slit experiments\cite{double}, tunneling of particles through barriers\cite{barrier}, Stern-Gerlach experiments\cite{stern} and the Aharonov-Bohm effect\cite{abeffect}. 

However, in the Bohmian approach quantum entanglement has not yet been analyzed quantitatively, although since Bell's construction of a hidden variable model for a single spin-1/2 system {(qubit)} \cite{bell66,mermin93} the formal Bohm approach was consistently extended to a causal theory capable of giving insight into Einstein-Podolsky-Rosen spin correlations in terms of well-defined individual particle trajectories with continuously variable spin vectors\cite{dewdney}. Today spin-1/2 systems can be treated causally in a nonrelativistic formalism based on the Pauli equation\cite{pauli}, by rigid rotor theory\cite{holland88},
{the hypersurface Bohm-Dirac model for entangled particles \cite{dur99}}
or by the Clifford algebra approach to Schr{\" o}dinger and Dirac particles \cite{hiley1}. 

Among the simplest quantum systems is a pair of  qubits for which quantum entanglement can be quantified, for example, by  
the entanglement of formation $E_F$ \cite{bennett}, the asymptotic conversion rate to maximally entangled states from an ensemble of  copies of a non-maximally entangled state \cite{vedral}. The entanglement of formation can be related to an associated quantity, concurrence $C$ \cite{wootters}. 

Here we consider  a qubit pair in a pure state
\begin{equation}
\left | \Psi \right \rangle=\cos { \vartheta \over 2} \left |\uparrow\downarrow\right \rangle+e^{i \varphi}\sin {\vartheta \over 2} \left |\downarrow\uparrow\right \rangle,
\label{psi}
\end{equation}
and for convenience we use spin-1/2 notation where $\left |\uparrow\downarrow\right\rangle$ corresponds to the first qubit in the "up" state, {\it i.e.}, in the direction of the $z$-axis, and the second qubit in the "down" state. 
The generalization to more general forms of qubit pairs including mixed states is possible, but is not considered here. Qubits are not restricted to real electron spins, but may be realized by any two state quantum system such as for example, entangled photon\cite{fotoni},  flux qubit in a superconducting ring\cite{makhlin}, charge pseudo-spin of electron pairs in a double quantum dot\cite{mravlje},  flying qubits in quantum point contacts \cite{rejec} or qubits in a composite system\cite{delfot}.

{{\bf De Broglie-Bohm formalism for spin-1/2.} --}
In order to demonstrate how quantum entanglement can be described ontologically we closely follow the approach introduced by Holland\cite{holland88,holland00} where the starting point is the mapping between a quantum rigid rotor and a classical spinning top in the presence of a quantum potential. 
The orientation of a rigid rotor is expressed by Euler angles $\zeta=\{\alpha,\beta,\gamma\}$  and, in the quantum approach, by the wave function $\psi(\zeta)$. 

Defining a differential operator $\hat{\mathbf{M}}$, whose components are the infinitesimal generators of the rotation group $SO(3)$, the quantum Hamiltonian is given by ${\hat{H}}=\hat{\mathbf{M}}^2/2 I$, where $I$ is an auxiliary parameter ("moment of inertia"), with $I\to0$ in the final results. {
The wave function is expressed as $\psi=\mathcal{R} e^{i \mathcal{S}}$, where $\mathcal{R}(\zeta)$ and $\mathcal{S}(\zeta)$ are real functions. Bohmian space angular momentum is then given by a real three dimensional vector ${\mathbf{M}}=i{\hat{\mathbf{M}}} \mathcal{S}$. This relation is an analogue of a more familiar de Broglie's guidance equation for the velocity  of a point-like particle with mass $m$ treated in the Bohmian approach, $m {\mathbf{v}}=\nabla \mathcal{S}$ \cite{bohm1}.}

The dynamics is determined from a Hamilton-Jacobi-type equation corresponding to the classical Hamiltonian
\begin{equation}
H={{\mathbf{M}}^2 \over 2 I}+Q,\quad Q={\hat{\mathbf{M}}^2 \mathcal{R}\over 2I \mathcal{R}},
\label{hclass}
\end{equation}
{where the quantum potential $Q$ generates  a quantum torque $\mathbf{T}=-i \hat{\mathbf{M}} Q$, which rotates the angular momentum vector via the equation of motion $d \mathbf{M}/dt=\mathbf{T}$ along the trajectory $\zeta(t)$. This is a 
counterpart  of the Newton equation for the case of a free particle in the Bohm formulation given by 
$m d{\mathbf{v}}/dt=\nabla({\nabla^2 \mathcal{R}/ 2m \mathcal{R}})$. The equation of the angular momentum motion simplifies to a set of first order non-linear differential equations where the solutions $\zeta(t)$ represent orbits in the configuration space, uniquely  determined by the initial positions $\zeta(0)$, and the angular momentum emerges as
${\mathbf{M}}[\zeta(t)]$. }
 \begin{figure}[htbp]
\begin{center}
\includegraphics[width=88mm]{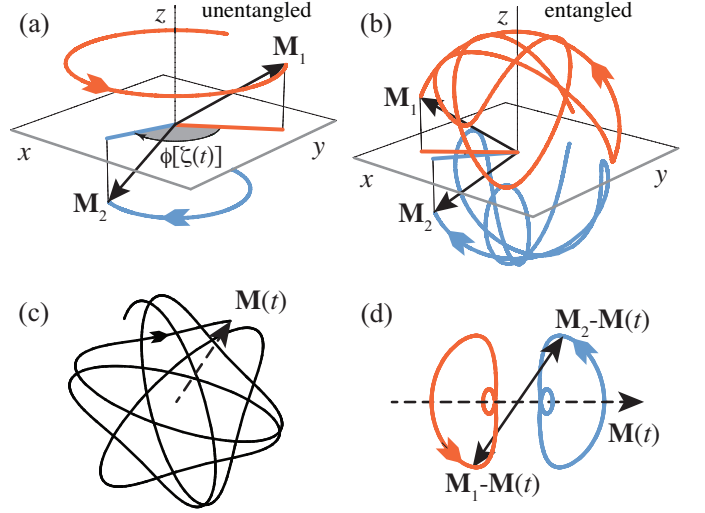}
\caption{ {(a)} An unentangled qubit pair -- momenta  {are independent} and precess  in opposite directions around the $z$-axis. {(b)} Typical motion of momenta in a partially entangled state,  $\vartheta=\pi/5$ and $\varphi=0$, with initial values for the  rotors $\zeta{(0)}=\{\alpha_1,\beta_1,\gamma_1,\alpha_2,\beta_2,\gamma_2\}=\{2,5,0,\frac{1}{2},0,0\}$. {(c)} Spirographic motion of ${\mathbf{M}}(t)={\mathbf{M}}_1+{\mathbf{M}}_2$ projected onto the $xy$-plane (parameters as in ({b}). {(d)} Trajectories of ${\mathbf{M}}_{1,2}$ shown in the $xy$-coordinate system rotating with ${\mathbf{M}(t)}$. In this coordinate frame the trajectories are closed and periodic with the period $\tau(\zeta)$.}
\label{Fig1}\vskip .5 cm
\end{center}
\end{figure}

Solutions for a single spin-1/2 rotor are known.
Quantum basis eigenstates of $\hat{\mathbf{M}}$ belong to the spin-1/2 sector whereas the Bohmian description embodies the coexistence of the motion of a rigid top, whose configuration space is $SO(3)$ and a guiding wave, whose spin configuration space is $SU(2)$ \cite{holland88}. {The pilot wave is
given in terms of the Wigner $D$-matrices. For the "spin up" state, for example, 
$\mathcal{R}_\uparrow(\zeta)\propto \cos {\alpha \over 2} $ and $\mathcal{S}_\uparrow(\zeta)=-(\beta +\gamma)/2$.}

From the equations of the motion follows a time dependent angular momentum ${\mathbf{M}}$: "spin up" and "spin down" means that ${\mathbf{M}}$, starting from some initial direction which is a function of the initial value $\zeta(0)$, precesses around the $z$-axis counterclockwise and clockwise, respectively, with constant $z$-axis projection ${\mathbf{M}}_{z}=\pm\frac{1}{2}$. {The angular  momentum magnitude is constant and can reach any value $|{\mathbf{M}}|\geq\frac{1}{2}$, depending
on the initial choice of $\zeta(t)$.}
Averaging over the initial values yields a {quantum equilibrium}  ensemble averaged angular momentum $\langle{\mathbf{M}}\rangle_{\mathrm{B}}$, which is time independent and identical to the quantum mechanical expectation value of the spin operator $\langle{\hat{\mathbf{S}}}\rangle$ and also to the time average of Bohm angular momentum $\langle{\mathbf{M}}\rangle_T$ for each particular ensemble representative $\zeta$. 
{In Fig.~1(a) is shown an example 
of two unentangled qubits in the state $\left |\uparrow\downarrow\right \rangle$. The solution is an independent motion of "spin up" and "spin down" vectors ${\mathbf{M}}_1$ and ${\mathbf{M}}_2$.

For a 
two qubit state  is the guiding function $\psi(\zeta)$ expressed by six variables $\zeta=\{\zeta_1,\zeta_2\}$. It should be noted that even for non-interacting, but entangled qubits the corresponding equations of the motion are coupled by the quantum potential $Q(\zeta)$ which incorporates their interaction in the corresponding Bohmian two-particle Hamiltonian. The solutions for each of the 
time-dependent angular momentum vectors ${\mathbf{M}}_{1,2}$  are determined by six common initial values $\zeta(0)$. }

{{\bf The probability distributions and the quantum entanglement.} --}
Let us now discuss results for a  qubit pair in the state parametrized by equation (\ref{psi}). 
The total angular momentum projection ${{M}}_{1z}+{{M}}_{2z}$ is zero while {
the angular momenta due to the action of the non-local quantum potential ({\it i.e.}, interaction) and the corresponding quantum torques exhibit a complex precessional motion, as illustrated for a particular choice of $\zeta(0)$ in Fig.~1(b). 
In general, only a few special cases have been explored so far. For example,}
for a fully entangled singlet state, $\vartheta=\pi/2$ and $\varphi=\pi$, the total momentum ${\mathbf{M}}={\mathbf{M}}_1+{\mathbf{M}}_2$ vanishes for each $\zeta$ in accordance with the usual imagery\cite{holland00}. 

{In this letter we concentrate on dynamical properties of the angular momenta relevant to the analysis of quantum entanglement while various other spin-spin correlation functions  and the corresponding probability distributions will be presented elsewhere \cite{ramsak3}.
We computed} trajectories ${\mathbf{M}}_{1,2}[\zeta(t)]$  covering the full configuration space with $\sim10^6$ initial values  $\zeta{(0)}$ per $| \Psi\rangle$, {{\it i.e.}, for a particular choice of $\vartheta$ and $\varphi$}. Although these trajectories exhibit extremely rich variety, some common properties can be outlined. (i) The quasi-periodic motion appears chaotic and, except in special cases, the projections of the total momentum ${\mathbf{M}}$ onto the $xy$-plane winds around the origin an infinite number of times in a spirographic manner\cite{astronom}, forming a  dense annulus limited by fixed outer and inner radii  [Fig.~1(c)]. (ii) The curve corresponding to relative momentum  ${\mathbf{M}}_2-{\mathbf{M}}_1$  is closed and periodic if plotted in the reference frame rotating  synchronously with ${\mathbf{M}}$ around the $z$-axis [Fig.1(d)].

{These results clearly show  that the} entanglement properties of such a qubit pair are reflected in the dynamics of the azimuthal angles $\phi_1{[\zeta(t)]}$ and $\phi_2{[\zeta(t)]}$ of angular momenta as follows. First, the ensemble average difference of azimuthal angles  $\phi[\zeta(t)]=\phi_2-\phi_1$ is time independent and given by $\langle \phi \rangle_{\mathrm{B}}=\varphi$, where the  average is defined by $\langle f \rangle_B=\int f(\zeta)\mathcal{R}^2(\zeta) {\mathrm d} \zeta$, with
${\mathrm d}\zeta=\prod_{i=1}^{2} \sin  \alpha_i d \alpha_i d\beta_i d\gamma_i$. The  corresponding probability distribution 
{
$dP(\phi)/ d  \phi=\int\delta[\phi-\phi(\zeta)]\mathcal{R}^2(\zeta) {\mathrm d} \zeta$ \cite{numerika}}
is constant for unentangled qubits and becomes progressively  peaked at $\varphi$ for increasing entanglement,
as presented in Fig.~2, culminating in precession of angular momenta at equal relative angle $\phi[\zeta(t)]=\varphi$ for all $\zeta$ consistent with perfect entanglement. {The shape of the distribution is independent of $\varphi$.}
 \begin{figure}
\begin{center}
\includegraphics[width=65mm]{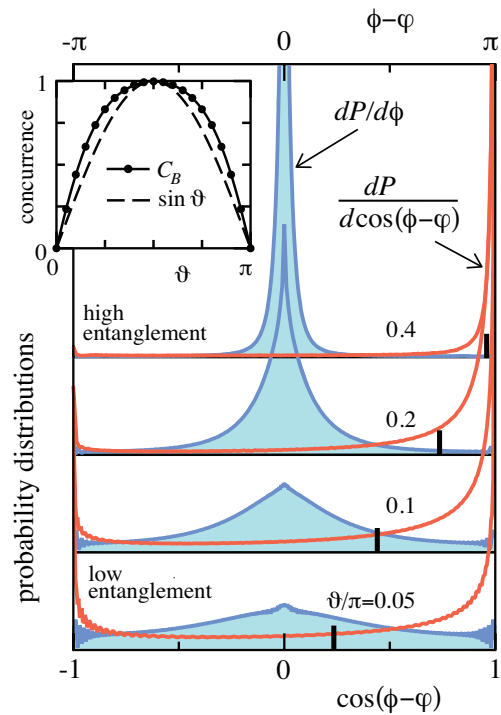}
\caption{ Probability distribution for relative angle $\phi$ of the in-plane projections of momenta (see Fig.~1(a)), $dP(\phi)/ d  \phi$ (blue, upper scale) shown relative to the average angle $\langle \phi \rangle_B=\varphi$ for various degrees of entanglement.  In the low entanglement regime the distribution is flat, becoming peaked at $\varphi$ in the fully entangled state. The corresponding probability distributions $dP(\cos(\phi{-\varphi}))/ d  \cos(\phi{-\varphi})$ (red, lower scale). The average of 
$\langle\cos(\phi{-\varphi})\rangle_B=C_B(\vartheta)$ is marked with thick vertical lines and shown also in the inset with bullets together with concurrence $\sin \vartheta$ (dashed line). Small oscillations are due to  finite size effects.
}
\label{Fig2}
\end{center}
\end{figure}

The next quantity of interest is the probability distribution 
$dP(\cos(\phi{-\varphi}))/ d  \cos(\phi{-\varphi})$, also presented in Fig.~2 and, in particularly 
the average cosine  which appears to be an excellent measure of entanglement, ranging from zero for an unentangled state to unity for a maximally entangled state. 
One can {readily} prove the exact expressions $\langle\cos \phi \rangle_\mathrm{B} =C_B \cos \varphi$ and $\langle\sin \phi \rangle_\mathrm{B} =C_B \sin \varphi$ or, equivalently,
\begin{eqnarray}
\varphi&=&\langle \phi \rangle_B,\\
C_B&=&\langle\cos (\phi-\varphi) \rangle_B= \\ 
&=&\sqrt{\langle\cos \phi \rangle_B ^2 +\langle\sin \phi  \rangle_B^2},\\
1-C_B^2&=&(\Delta \cos \phi)^2+(\Delta \sin \phi)^2,
\label{c}
\end{eqnarray}
where {$C_B$ is dependent only of $\vartheta$} and $\Delta \cos \phi$, $\Delta \sin \phi$  are standard deviations from average cosine and sine, respectively. 
{Note that $\langle \sin (\phi-\varphi)\rangle_B=0$.} 
Similar formulae can be derived also by appropriately defined cosine and sine operators in standard quantum mechanics formalism  \cite{ramsakQM}.
A higher degree of entanglement can thus be visualized as a highly correlated distribution of angular momenta making azimuthal angles {difference} close to $\varphi$, with suppressed fluctuations for progressively increasing entanglement.

In the Bohmian picture of entangled qubit pairs the quantity $C_B(\vartheta)$ (inset to Fig.~2) plays the role of Wootters concurrence given by $C{(\vartheta)}=2 |\langle\Psi| \hat{S}^+_1 \hat{S}^-_2 |\Psi \rangle|=\left|\sin \vartheta\right|$ \cite{wootters}, where $\hat{S}^\pm_{1,2}$ are spin-ladder operators for qubits 1 and 2, respectively\cite{amico08,ramsak1}.
High  $C_B(\vartheta)\to1$ signals weak angle fluctuations whereas for $C_B(\vartheta)\to0$, entanglement is suppressed and progresively destroyed when the standard deviation of the cosine (or sine) is comparable to its average. 

Concurrence is related to quantum mechanical expectation values and $C_B$ to ensemble averages. A natural question arises: Is there some imprint of entanglement, not only in the full ensemble, but also in each particular representative $\zeta$?  Standard quantum mechanics does not discuss such questions, while in the Bohmian approach one can classify the ensemble further.

Each representative $\zeta$ will in general lead to a different time average of $\cos \phi([\zeta(t)]{-\varphi})$, defined by
\begin{equation}
\langle \cos \phi\rangle_T= {1 \over \tau}\int_0^{\tau} \cos (\phi[\zeta(t)]{-\varphi})  {d t },
\end{equation}
where $ {\tau}(\zeta)$ is the period corresponding to the trajectory $\zeta(t)$. Time averaged $\cos  (\phi[\zeta(t)]{-\varphi})$ is distributed according to the
probability distribution $dP(\langle \cos\phi \rangle_T)/ d \langle \cos\phi \rangle_T$.  In Fig.~3 is presented the distribution for several representative values of $\vartheta$. The distributions  
in the strong entanglement regime, $\vartheta \sim \pi/2$, are peaked at $\langle \cos \phi\rangle_T\sim 1$,
similar to results in Fig.~2 but strikingly different in the weak entanglement regime. 
This is understandable since for  weakly entangled qubit pairs the angular momenta precess almost independently with a vanishing time average $\cos (\phi-\varphi)$ for every $\zeta$, leading to the zero-peak in the distribution. This is contrary to the fixed time distribution which is flat because of the randomly distributed angles.
 \begin{figure}
\begin{center}
\includegraphics[width=75mm]{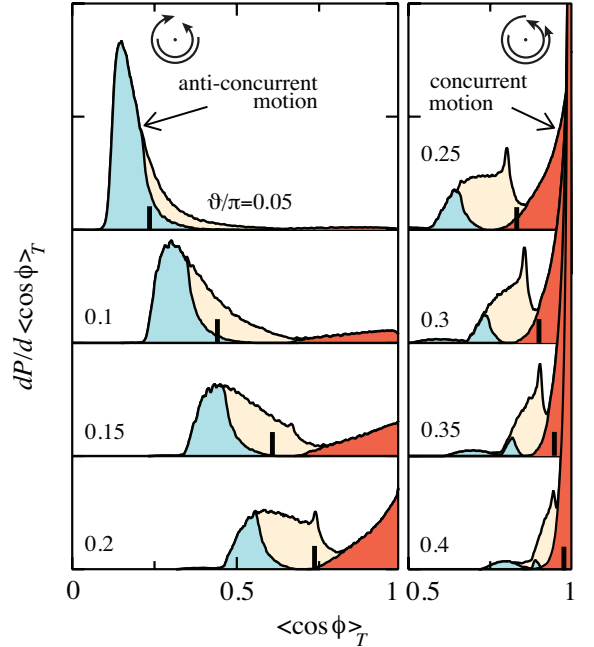}
\caption{ Time averaged $\cos(\phi{-\varphi})$ probability distributions for various degrees of qubit pair entanglement. 
The concurrent {($\mathcal{C}_\zeta=1$)} and anti-concurrent {($\mathcal{C}_\zeta=-1$)} fractions are indicated by red and blue shading, respectively. Thick vertical lines represent distribution averages  (values are identical to those in Fig.~2,
{{\it i.e.}, $C_B$}).}
\label{Fig3}
\end{center}
\end{figure}

It should be noted that 
the ensemble representing two qubits is non-ergodic, {\it i.e.}, ensemble averages $\langle f \rangle_\mathrm{B}$ do not generally equal time averages $\langle f \rangle_T$. For the present case of noninteracting (but entangled) qubits 
$\langle f \rangle_B$ is independent of time (as it should be) and $\langle \langle f \rangle_T \rangle_B=\langle f \rangle_B$, { where the Bohmian ensemble average of $\langle f \rangle_T$ is evaluated using the corresponding initial values $\zeta(0)$}. 

The probability distribution for the cosine time average exhibits a distinctive tripartite structure. We found that the discriminating property of ensemble representatives  is the relative direction of angular momenta precession. In the low entanglement regime the $xy$-plane projection of momenta ${\mathbf{M}}_1$ and ${\mathbf{M}}_2$ precess mainly in opposite directions. An extreme case is an unentangled state, Fig.~1(a). In general, momentum pairs move part time in the same and part time in the opposite direction. We classify representatives that {\it always} precess in the same direction as  "concurrent" movers whereas those which always precess in the opposite direction are classified as "anti-concurrent". 

{
To be specific, we introduce a measure $\mathcal{C}_\zeta$ to distinguish different trajectories $\zeta(t)$ according to their "concurrency",
\begin{equation}
\mathcal{C}_\zeta
= {1 \over \tau}\int_0^\tau \mathrm{sign}{d \phi_1[\zeta(t)]\over dt}{d \phi_2[\zeta(t)] \over dt}  {d t }.
\label{concurrency}
\end{equation}
At each moment the angular momenta for a given trajectory $\zeta(t)$ precess either in the same or in the opposite direction. One can thus visualize the concurrency as a measure of the share of the time that both angular momenta move in the same direction. For example, $\mathcal{C}_\zeta=\pm1$ for perfectly concurrent and anti-concurrent movers, respectively, and $\mathcal{C}_\zeta>0$ for  trajectories where angular momenta move concurrently more than half of the time for some members of the ensemble.
}
 \begin{figure}[tbp]
\begin{center}
\includegraphics[width=80mm]{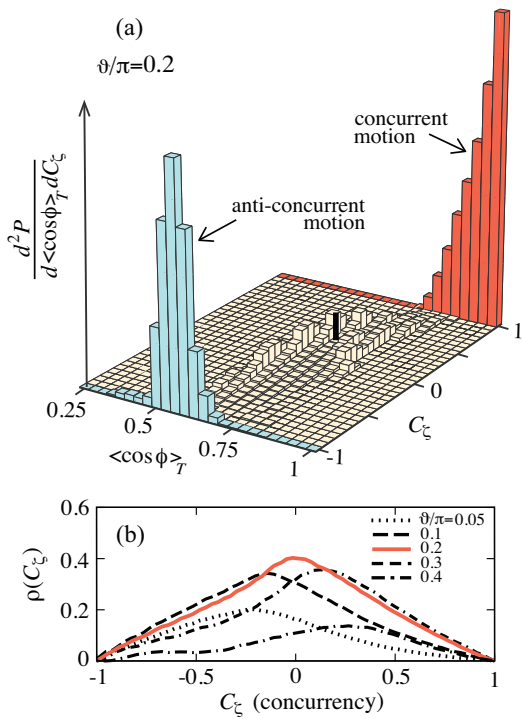}
\caption{{(a) The histogram $d^2 P/d \langle \cos \phi\rangle_T d\mathcal{C}_\zeta$ extracted from $\sim10^6$ trajectories for $\vartheta=\pi/5$ \cite{numerika}. The samples are identical to those in Fig.~2, while a larger bin size $\frac{1}{16}$ is applied here in order to enable the presentation of a discrete/continuous type of the distribution in a single plot. Thick vertical line represents the average point \{$C_B $, $\langle \mathcal{C}_\zeta\rangle_B$\}. (b) Continuous part $\rho(\mathcal{C}_\zeta)$ vs.  concurrency for various $\vartheta$. The distribution  Eq.~(\ref{dpdcon}) is normalized, thus  
$\int_{-1}^1  \rho(\mathcal{C}_\zeta) d\mathcal{C}_\zeta=1-(P_+ +P_-)$.}}
\label{Fig4}
\end{center}
\end{figure}

This is shown in Fig.~3  where distributions for concurrent and anti-concurrent movers are shaded red and blue, respectively. In the low-entanglement regime anti-concurrent movers dominate whereas the distribution  of concurrent movers progressively dominates as entanglement increases. Remarkably, only a minority  of representatives move both concurrently and anti-concurrently, part of the time
{($|\mathcal{C}_\zeta|<1$). 

Concurrency should not be confused with concurrence which is given by 
the ensemble average of $\cos\phi$ while, on the other hand, concurrency relates to spin-precession of a particular pair in the Bohmian ensemble and has no direct quantum analogue.
The probability distribution of concurrency consists of a discrete and a continuum part,
\begin{equation}
{d P(\mathcal{C}_\zeta)\over d\mathcal{C}_\zeta}=P_+\delta(\mathcal{C}_\zeta-1)+
P_-\delta(\mathcal{C}_\zeta+1)+\rho(\mathcal{C}_\zeta),
\label{dpdcon}
\end{equation}
where $P_{\pm}$ is the probability that the concurrency 
is exactly 
$\pm1$, respectively, and 
$\rho(\mathcal{C}_\zeta)$ is a continuous function  \cite{delta}, shown for various $\vartheta$ in Fig.~4(b), for which motion is
sometimes concurrent and sometimes anti-concurrent as $t$ changes. Hence the probability density $\rho(\mathcal{C}_\zeta)$ tends to zero as  $|\mathcal{C}_\zeta|\to1$, which strictly separates (anti-)concurrent motions from the partially-concurrent regime. 
For a typical case of a partially entangled qubit pair with $\vartheta=\pi/5$ is in Fig.~4(a) shown 
the histogram probability distribution $d^2 P/d \langle \cos \phi\rangle_T d\mathcal{C}_\zeta$. Coarse binning is applied in order to emphasize the middle regime $|\mathcal{C}_\zeta|<1$ together with  
(anti-) concurrent parts.

Let us emphasize, that}
concurrent motion is inherent to an individual representative and their share $P_+$ in the ensemble can be considered as a suitable measure of entanglement for Bohmian ensembles. 
Such a clear visualization of entanglement cannot be deduced from quantum mechanics where only averages of operators are accessible. 
Quantifying entanglement is important in applications which require maximally entangled (Bell) pairs to be shared by two remote parties. In reality  qubit pairs in general are not perfect Bell states and  entanglement distillation is required in order to
extract such pairs\cite{bennett}. 
The entanglement of formation $E_F$ is an upper bound to the average number of Bell pairs that can be extracted or distilled from a set of copies of an entangled state, using only local operations and classical communication. In the Bohmian approach the probability  for concurrent motion $P_+$ may be considered a counterpart of $E_F$. This probability is shown in Fig.~5(a) together with $E_F$ (dashed line). The probability for anti-concurrent motion $P_-$ corresponds to the region above the $1-P_-$ line.

 \begin{figure}[tbp]
\begin{center}
\includegraphics[width=52mm]{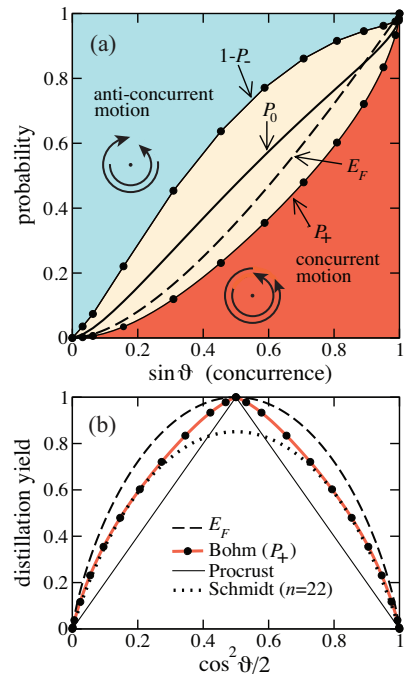}
\caption{(a) $P_+$ line represents the probability that momenta move concurrently (shaded red), and above $1-P_-$ is the probability that motion is anti-concurrent (blue shaded region). The $P_0$ line denotes a
share of trajectories which move concurrently more than half of the time {($P_0=P_++\int_{0}^1  \rho(\mathcal{C}_\zeta) d\mathcal{C}_\zeta$)}.
Entanglement of formation $E_F$ is shown with a dashed line.  (b) Entanglement of formation vs. the probability for $\left |\uparrow \downarrow\right\rangle$ configuration in $|\Psi\rangle$ (dashed). Bullets represent the probability for concurrent motion -- the Bohmian counterpart of $E_F$. The full line represents the distillation yield using Procrustean method. The dotted line is the yield using the Schmidt projection method with $22$ input pairs.  } 
\label{Fig5}
\end{center}
\end{figure}

In Fig.~5(b) the distillation yield of Bell pairs obtained by the Procrustean method\cite{bennett} is plotted as a function of $\cos^2 \vartheta/2$  for one qubit pair (straight full line) together with the upper limit, the entanglement of formation (dashed line). The probability for concurrent movers in the Bohmian picture (red line) is above both the Procrustean method and the yield of Schmidt projection methods when applied with $n=22$ input pairs\cite{bennett}. This result suggests that in the Bohmian picture  "preformed entangled" pairs exist and their extraction would represent a quantum distillation protocol with high yields obtained from a single qubit pair.

{{\bf Conclusion.} -- For the sake of completeness let us discuss also the class of states
\begin{equation}
|\widetilde{ \Psi}\rangle=\cos { \vartheta \over 2}\left |\uparrow\uparrow\right\rangle+e^{i \varphi}\sin {\vartheta \over 2} \left|\downarrow\downarrow\right\rangle.
\label{tpsi2}
\end{equation}
%
The formalism can be applied in a similar manner as for the case of $\left |\Psi\right\rangle$ while the main distinction is that for states  Eq.~(\ref{tpsi2})  the sum of azimuthal angles, $\tilde \phi=\phi_1+\phi_2$,  is important, not the difference $\phi$. Although the trajectories with equal initial values, but different quantum states $|\widetilde{ \Psi}\rangle$  ($|\Psi\rangle$), are altogether different,
the probability distributions   take  the same form after the appropriate substitution  
$\phi \leftrightarrow\tilde\phi$. For example, $\langle \tilde\phi \rangle_B=\varphi$ and, furthermore, 
the sign of concurrency is reversed, {\it i.e.}, $\widetilde{\mathcal{C}}_\zeta=-\mathcal{C}_\zeta$.}

In brief summary the main findings are as follows. In the usual quantum mechanical approach, two main measures are used to quantify entanglement, namely concurrence and the entanglement of formation. These rather abstract concepts are related but rather difficult, if not impossible, to visualise. Conversely, in the Bohmian approach it is possible  to define analogous measures that are quantitatively similar (though not identical) to concurrence and entanglement of formation but are quite distinct from each other and may be directly related to intuitively appealing geometric interpretations. In the Bohmian interpretation, the angular momentum vectors of the two particles (qubits) precess in a well-defined way with some initial probability distribution. The Bohmian concurrence is simply the average mutual cosine of the $xy$-plane projection of the two angular momentum vectors and is a property of the whole ensemble. In standard quantum mechanics this corresponds to an expectation value. On the other hand, the Bohmian entanglement of formation {depends on a property --
concurrency -- } of each representative of the ensemble which either has "concurrent motion" (the angular momentum vectors precess in unison) or they do not. The share of representatives with this motion is the entanglement of formation. Apart from this simple and appealing underlying picture, it may also have practical relevance since it suggests that distillation protocols could have a very high yield already for one single qubit pair. This ontological result is entirely missed (has no meaning) in the usual quantum mechanical approach. {Our results also give rise to a challenging question:  Can some quantity identical, or analogous, to concurrency also manifest itself in other 
spin-1/2 formalisms?}

The author thanks T. Rejec, I. Sega and T. Huljev {\v C}ade{\v z}  for discussions, J. H. Jefferson for valuable suggestions
and C. Dewdney, B. J. Hiley and P. R. Holland for clarifying  some details of Bohmian spin-1/2 theories. He also acknowledges the support from the ARRS  under Contracts No. J1-0747 and P1-0044.S

\vfill

\begin{thebibliography}{30}
\bibitem{numerika}{For the results presented throughout  this paper 
we used a uniform grid of  $128^3$ points in the configuration space of $\{\cos\alpha_1,\cos\alpha_2,\beta_2-\beta_1\}$.  
Hidden variables $\gamma_{1,2}$ corresponding to the internal rotation around the body principal axis are irrelevant for the quantities considered here and the Dirac delta-function was  implemented as a normalized rectangular function of width $1/100$.}
\bibitem{delta} {The support of the concurrency  is restricted to the interval $|\mathcal{C}_\zeta|\leq1$, therefore the relation $\delta(\mathcal{C}_\zeta\pm1)$ is to be read as $\lim_{\epsilon\to0^+} \delta[\mathcal{C}_\zeta\pm(1-\epsilon)]$.}
\end{thebibliography}
\end{document}